# Systematic approach to the growth of high-quality single-crystals of $Sr_3Ru_2O_7$

**Authors**

R. S. Perry[1,2] and Y. Maeno[1,2]

[1] Department of Physics, Kyoto University, Kyoto 606-8502, Japan.
[2] Kyoto University International Innovation Center, Kyoto 606-8502, Japan.

**Abstract**

We describe a simple procedure for optimising the growth condition for high quality single crystals of the strontium ruthenate perovskites using an image furnace. The procedure involves carefully measuring the mass lost during crystal growth in order to predict the optimal initial atomic ratio. Using this approach we have succeeded in growing crystals of $Sr_3Ru_2O_7$ with a residual resistivity as low as 0.25 µΩcm. The procedure we describe here is expected to be useful for other systems when a standard travelling-solvent floating-zone (TSFZ) method cannot be used because of high volatility of a constituent material.

**1. Introduction.**

The strontium ruthenates series of materials are proving to be a fascinating area for investigating strongly correlated electron behaviour. In particular, the single layered $Sr_2RuO_4$ exhibits unconventional spin-triplet superconductivity that condenses out of a highly correlated metallic state [1,2]. More recently, the bilayered member $Sr_3Ru_2O_7$ has been shown to be an itinerant metamagnet that is close to a quantum critical point [3,4]. An important issue in the ruthenates is crystal purity [5,6]. Most notably, the superconductivity in $Sr_2RuO_4$ crystals survives only for residual resistivities $\rho_{res}$ of less than 1.0 µΩcm [7]. The crystal growth of $Sr_2RuO_4$ has been well documented and single crystals with $\rho_{res}$ ~0.1 µΩcm can be grown with minimal difficulty [8]. However, other members of the Ruddlesden-Popper perovskite series are proving more difficult to synthesize in high-quality single crystal form. Single crystals of $Sr_3Ru_2O_7$ were first synthesised using a floating zone technique by Ikeda and co-workers [9] although their highest quality crystals had a modest $\rho_{res}$ ~ 2.0 µΩcm. The preparation and growth techniques of $Sr_2RuO_4$ and $Sr_3Ru_2O_7$ are almost identical and it is currently unclear as to why the single crystals of $Sr_3Ru_2O_7$ are an order of magnitude less pure than crystals of $Sr_2RuO_4$.

In this paper, we describe a systematic procedure to grow single crystals of $Sr_3Ru_2O_7$ with $\rho_{res}$ as low as 0.25 µΩcm using a floating zone technique. The starting point for this study was based on the work done by Ikeda and co-workers [9]. The goal of growth of stoichiometric crystals is often hampered by mass loss of one of the constituent components. The method used by Ikeda to circumvent this problem was to compensate for the mass loss by increasing the initial amount of the constituent that is lost. In this report, we term this the flux feeding floating zone technique (FFFZ). We have extended this technique by employing a simple procedure involving measuring the mass of material evaporated during crystal growth. This allows the atomic ratio of ruthenium to strontium in the synthesized crystal to be estimated. Crystals with a non-stoichiometric atomic ratio are inherently disordered and this is can be the cause of low quality crystals grown using an image furnace. Our technique allows us to predict the optimal nominal ratio of components to grow stoichiometric crystals. We also investigate how the evaporation rate is affected by the growth variables of an image furnace. Finally, we discuss the nature and levels of impurity phases in the single crystals of $Sr_3Ru_2O_7$.

**2. The FFFZ technique**

Before we describe the crystal growth of $Sr_3Ru_2O_7$, it is useful to first discuss the floating zone crystal growth technique and its application to the crystal growth of different materials. In particular, growth of transition metal oxide using an image furnace is an increasingly popular technique. Briefly, radiation from halogen bulbs focused by ellipsoidal mirrors is used to melt the tip of a rod of polycrystalline material. The single crystal is then pulled continuously from the melt while the polycrystalline rod is continuously lowered into the melt to replenish lost material. Thus the molten zone is observed to 'float' between the poly- and single crystal rods. The rods are rotated counterclockwise during the crystal growth to mix the melt, and are encased inside a quartz tube to allow a variable atmosphere and pressure.

The most notable successes have been the crystal growth of the cuprate high temperature superconductors by Tanaka and Kojima [10]. They used a modified floating zone technique called the travelling-solvent floating-zone technique (TSFZ) [11]. This involves an initial flux rich in the transition metal oxide (e.g. CuO), which travels along the stoichiometric feed rod. This is possible because of the low volatility of the constituent compounds. However, this technique is less useful when one of the constituents has a high volatility at the temperature of coexisting melt and the target crystal. In some cases, the loss of the material from the melt can be so great that the molten zone moves past the peritectic point causing stabilization of undesired phases. To combat this an alternative technique called the FFFZ technique has been developed over the past decade [8,12]. A feed rod that is rich in the constituent with the high volatility is used instead of a stoichiometric feed rod. The excess constituent firstly acts as a flux for the floating zone and secondly, the evaporation is compensated for by the extra material from the feed rod. However, this method is not without its problems, most notably the creation of a new growth parameter: the amount excess material needed to be added to the feed rod to ensure a stoichiometric crystal is grown.

The single crystals of the strontium ruthenates were grown using an image furnace employing a Ru self-flux technique. We followed a similar preparation method described by Mao *et al.*, the precise details of which can be found elsewhere [8]. The starting materials were $SrCO_3$ (99.9% purity; Ba<7 ppm) and $RuO_2$ (99.99% purity) which were ground together in a mortar and pestle. The nominal ratio of the components is denoted as $n=2N(Ru)/N(Sr)$, where $N(Ru)$ and $N(Sr)$ are the molar numbers. The ground materials were then compressed into pellets (1 cm in diameter) and calcined at 1000°C in air for 18 hours. The polycrystalline material was then re-ground, pressed into a rod of 10 cm in length by 5 mm in diameter, and sintered at 1350°C in air for 2 hours to harden the rod. $CO_2$ is completely evaporated at this stage, but the loss of $RuO_2$ is negligible compared with the amount lost during the crystal growth. During calcination and sintering, care must be taken to avoid any contact between the materials and the crucible or the boat [8]. Finally, the rod was set into a two-mirror infrared image-furnace using 2 kW halogen lamps with flat shape filaments (NEC Machinery, model SC-K15HD-HP).

**2.1 Definition of the primary parameters**

It has been pointed out that image furnaces have several control variables that can alter the product crystal [13]. These can be broken down into two types of parameters: primary and secondary. Primary variables include the atmosphere content, the pressure $P$ and the crystal growth speed $V_1$. They are set at the start of the growth and usually remain unchanged for the duration of the growth. They also largely determine the nature of the crystal. The nominal molar ratio of the constituent compounds is also a primary variable in the FFFZ method and it needs to be optimized so the grown single crystal is stoichiometric. We shall return to this later in the paper. Secondary variables include the feed speed of the ceramic rod $V_2$, the power into the halogen lamps and the rotation speed of rods $\omega$ and are altered during the growth to obtain a stable molten zone. A stable molten zone occurs when the rate at which material enters the molten zone from the feed rod is equal to sum of the rate at which material leaves the zone as a crystal and the rate at which material is lost due to evaporation. In order to grow large single crystals, one is required to keep the floating zone stable by adjusting the secondary variables over a time period that is related to the growth speed. So, to understand how to improve the growth of $Sr_3Ru_2O_7$ it is useful to understand how the numerous image furnace parameters affect the crystal produced.

**2.2 Evaluation of the evaporated $RuO_2$ during the growth**

As we mentioned earlier, the high volatility of one of the constituents is a common problem during crystal growth using an image furnace [14]. In the strontium ruthenates, $RuO_2$ has a high vapour pressure. Mao *et al.* used the FFFZ method to grow high quality single crystals of $Sr_2RuO_4$ [8] and they empirically determined an optimal $n$ of 1.15 (stoichiometric $n$ for $Sr_2RuO_4$ is 1.00). This was discovered by trial and error and, to the best of our knowledge, little has been done to quantify the magnitude of the evaporation and its dependence on other parameters of the crystal growth. Indeed, $n$ and $V_1$ were the main parameters Mao and co-workers used to optimise the crystal growth of $Sr_2RuO_4$. For the crystal growth of $Sr_3Ru_2O_7$ Ikeda *et al.* used $n=1.50$ (stoichiometric $n$ for $Sr_3Ru_2O_7$ is 1.33) [15]. They also used an increased pressure and atmospheric oxygen content not only to suppress the evaporation of $RuO_2$ but also to avoid formation of the eutectic of $Sr_2RuO_4$ and Ru [16]; however, the crystals had more than an order of magnitude lower quality in terms of $\rho_{res}$ compared to Mao's $Sr_2RuO_4$ crystals. It was suggested that the lower quality maybe due to ruthenium vacancies so in order to grow high quality $Sr_3Ru_2O_7$ information of the mass loss of $RuO_2$ is necessary.

If we assume that *only* $RuO_2$ is lost during the growth, we can write the equation: $Sr_2Ru_nO_y \rightarrow Sr_2Ru_{n'}O_{y'} + (n-n')RuO_2$ where $n'$ is twice the average ratio of Ru to Sr for the grown single crystal. The initial assumption can be checked by performing powder X-ray diffraction analysis on the evaporated material that collects on the inside of the furnace. The left-hand side of the equation is simply the polycrystal feed rod of mass $m_{rod}$ used during the growth and the right-hand side is the product crystal of mass $m_{cry}$ plus the total evaporated $RuO_2$ during the growth $m_{evap}$. $m_{rod}$ can be evaluated by measuring the mass of the polycrystalline rod before and after the growth and $m_{cry}$ is most accurately evaluated by measuring the mass of the detachable seed-crystal holder before and after the growth. The difference between $m_{rod}$ and $m_{cry}$ is then simply $m_{evap}$. The masses are related using the equation: $n' = n(1-L) - 1.557L$ where the mass loss, $L = m_{evap} / m_{rod}$. In this study, $m_{rod}$ is typically 8 g. It is important to note that $n'$ is only the *average* ratio over the whole crystal. If the crystal growth is unstable then many phases maybe stabilised throughout the crystal, which may not be reflected in $n'$. Still, knowledge of the mass loss $L$ can be used to predict the value of $n$ required to grow high-quality single-phase crystals of $Sr_3Ru_2O_7$ and can help to provide an insight into how the growth variables affect the mass lost.

We have repeated the crystal growths for $Sr_2RuO_4$ and $Sr_3Ru_2O_7$ using similar parameter to those given in references [8,15]. The growth conditions and analysis results are shown in the first two columns of Table 1. The residual resistivity $\rho_{res}$ is used as a measure of the quality of the single crystal. The resistivity was measured using a standard four probe technique on a Quantum Design PPMS system and $\rho_{res}$ was extracted by fitting the data to a Fermi liquid $T^2$ power law. Naturally, a larger $n$ is used to grow $Sr_3Ru_2O_7$ because it is richer in Ru than $Sr_2RuO_4$. Powder X-ray analysis was used to confirm that the crystals were single phase. We essentially reproduced the qualities of the crystals reported in references [8,15]. It is interesting to compare the quality of crystal grown with the resultant ratio $n'$. In the $Sr_2RuO_4$ growth, $n'=0.99$ indicating that the Sr and Ru are in approximately the correct stoichiometric ratio. However, the $Sr_3Ru_2O_7$ crystals have $n'=1.23$ suggesting that the $Sr_3Ru_2O_7$ crystals are ruthenium deficient. These observations are supported by the impurity resistivity $\rho_{res} \sim 0.2$ μΩcm and $\sim 2.0$ μΩcm for $Sr_2RuO_4$ and $Sr_3Ru_2O_7$, respectively. The materials and initial preparation method for both growths were identical implying that the low quality of the $Sr_3Ru_2O_7$ crystals is due to the growth conditions inside the image furnace and not due to extrinsic contaminants. Therefore, to produce high quality single crystals of $Sr_3Ru_2O_7$ we need to fine tune the crystal growth to synthesise single phase crystals with $n'=1.33$.

## 3. Optimisation of the primary parameters

We have varied the primary parameters $n$, atmospheric content, and $V_1$ and measured the mass loss $L$ to better understand how to optimise the crystal growth of the ruthenates.

### 3.1 The nominal Ru/Sr ratio $n$ versus $L$

Figure 1 demonstrates how mass loss $L$ increases with the concentration of Ru in the polycrystal feed rod. The mass loss $L$ has an approximately linear relationship to $n$, although the scatter on the data makes a quantitative relationship difficult to determine. Powder X-ray analysis on the evaporated material collected from the wall of the quartz tube confirmed that it was indeed $RuO_2$ that was evaporated. Each point on Fig.1 represents a stable growth run for which all of the control parameters except $n$ were kept constant. The requirement of a stable growth attempt is an important one. We define a stable growth as the achievement of a stable floating zone by altering the secondary parameters for a length of time that is at least 80% of total growth time. This allows an approximately constant evaporation rate to be maintained and a homogeneous crystal to be produced. Growths were made under several different atmospheres (all at 10 bar pressure) shown by different point symbols in Fig. 1.

### 3.2 Atmospheric content

Table 2 compares four crystal growths with a variable atmospheric content. $n$, $V_1$ and $P$ are constant for each growth. There is a systematic increase of $RuO_2$ mass loss with oxygen partial pressure $P_{O2}$. This indicates that the resultant ratio $n'$ drops, although the estimated absolute error of $L$ could be as large as ±1 % from the scatter on the data in Fig. 1. The decrease in $n'$ with increasing $P_{O2}$ suggests that the crystals grown under a high $P_{O2}$ should contain more fraction of $Sr_2RuO_4$. Crystals from each growth rod were also examined using powder X-ray analysis. This has an opposite conclusion: an increasing $Sr_3Ru_2O_7$ content with increasing oxygen percentage. This result may be attributable to Ru deficiencies in the crystals as well as other impurity phases. Thus we should be careful predicting the nature of the phase of a crystal solely from $n'$. The viscosity of liquid in the floating zones also decreases with increasing $P_{O2}$. This is demonstrated by the width of the crystal rod: low viscosity liquid flows easily downwards out of the molten zone to create a wide crystal. Interestingly, the temperature of the melt measured by a pyrometer is approximately constant, to within errors. The result is in agreement with Ikeda *et al.* [15] in that higher oxygen partial pressure does indeed help stabilise the three dimensional end of the strontium ruthenate series; however, there is an adverse effect of increasing the evaporation rate of ruthenium oxide. It should be noted that at very low oxygen partial pressure $Sr_2RuO_4$ and Ru metal are adjacent to the liquid phase but at higher partial pressure only the $Sr_2RuO_4$ and $Sr_3Ru_2O_7$ phases are stabilised.

### 3.3 The growth speed $V_1$

The growth speed is perhaps the most sensitive parameter of the crystal growth. The growth speed alters the diameter of crystal which changes the shape of the floating zone and hence the evaporation from it. Table 3 shows data from three crystal growth runs with different growth speed. Low growth speeds tend to stabilise the three dimensional perovskites whereas high growth speed is more conducive for $Sr_2RuO_4$ to grow. We determined the optimal growth speed of $Sr_3Ru_2O_7$ empirically to be around 15 mm/hr and have been unable to grow large single crystals of $Sr_3Ru_2O_7$ at any growth speed other than this. The data suggests that the optimal growth speed might coincide with a minimum in the mass loss but we cannot confirm this.

## 4. The optimised growth of $Sr_3Ru_2O_7$

### 4.1 Summary of the procedure to optimise the growth parameters

Given the information described in the previous section, let us summarize the procedure we used to grow high quality single crystals of $Sr_3Ru_2O_7$. For high quality single crystals of $Sr_3Ru_2O_7$, we desire $n'=1.33$ thus, in a sense, $n$ is a dependent variable. In order to determine the correct growth parameters we followed a simple procedure. Firstly, we optimized the static, thermodynamic parameters $P$ and $P_{O2}$. Knowledge of how $L$ depends on $P$ and $P_{O2}$ is useful here. In the ruthenates, for example, a high $P$ and intermediate $P_{O2}$ are required to minimize the evaporation of $RuO_2$. Secondly, we optimized the dynamic, primary parameter $V_1$. Finally, we use an iterative method to optimise $n$. This involves the measurement of $L$ after each growth, which allows us to predict a value of $n$ for the next growth that will bring us closer to our desired $n'$. Once crystals with $n'\sim1.33$ are produced, powder X-ray analysis is then used to confirm that the crystals were single phase. If the crystals were multi-phase, one of the primary variables was re-adjusted and the iterative procedure repeated. In our experience, $V_1$ is the most sensitive parameter, while atmospheric content and pressure can be used for fine tuning the growth to achieve higher quality crystals. Initially, we tried to grow $n'=1.33$ crystals in a 100% $O_2$ atmosphere. Although we achieved an $n'=1.33$ (with a large 13.8% mass loss), X-ray analysis showed that the crystals contained $Sr_2RuO_4$ and $Sr_4Ru_3O_{10}$ phases, in addition to $Sr_3Ru_2O_7$. This suggests that high-quality single-crystal growth of $Sr_3Ru_2O_7$ using a 100% $O_2$ atmosphere is difficult.

The growth parameters for $n'=1.33$ mono crystals of $Sr_3Ru_2O_7$ is shown in the third column in Table 1. For optimal growth, we used a rod of diameter 5 mm, $\omega = 45$ rpm and a $V_2\sim 14$ mm/hr. We have succeeded in synthesising crystals with $\rho_{res}$ as low as 0.25 $\mu\Omega$cm, which is comparable to the quality of the $Sr_2RuO_4$ crystals. We are also able to grow single crystals of up to 1 g mass, which are suitable even for neutron measurements (see Fig. 2). The main difference in the growth conditions between Ikeda's crystals and the present study is the $P_{O2}$ used. We used a 10%$O_2$+90%Ar atmosphere which acts to lower the evaporation of the $RuO_2$ from the melt. A lower mass loss is generally desirable because it makes the floating zone more stable. The viscosity of the liquid in the floating zone is also greater using a 10%$O_2$+90%Ar atmosphere. This effect is difficult to quantify but the result is that the crystal rod is thinner and more homogeneous. For example, the diameter of the crystal rod is typically ~7.5mm and ~3.8mm using 100% $O_2$ and 10%$O_2$+90%Ar atmosphere, respectively. In Fig. 3 we plot the low temperature resistivity of crystals grown under both 100% $O_2$ and 10%$O_2$+90%Ar atmospheres. The improvement in quality is clearly demonstrated.

**4.2 The evaluation of phase purity of our $Sr_3Ru_2O_7$ single crystals**

As mentioned earlier, the single-phase nature of the crystals were confirmed using powder X-ray diffraction as well as by inspection using a polarised light microscope and in most samples no alien phases were observed. However, in some samples $Sr_4Ru_3O_{10}$, $SrRuO_3$ and $Sr_2RuO_4$ were detected. $Sr_4Ru_3O_{10}$ and $SrRuO_3$ are both itinerant ferromagnets with Curie temperatures of ~90K and 160 K, respectively [17,18], while $Sr_3Ru_2O_7$ is paramagnetic. A SQUID magnetometer (Quantum Design MPMS system) was used to measure the residual ferromagnetic moment in the crystals. The results indicated ferromagnetic inclusions of typically 70 ppm in terms of molar ratio, although some crystals contained order of magnitude smaller inclusions. The presence of a $Sr_2RuO_4$ impurity phase is more difficult to detect. However, since $Sr_2RuO_4$ is a superconductor with a $T_c$ of 1.5 K, we would expect the superconducting $Sr_2RuO_4$ inclusions to short-circuit current paths through the bulk $Sr_3Ru_2O_7$. Indeed, some of the crystals exhibited a partial resistive drop attributable to $Sr_2RuO_4$ inclusions. However, no superconductivity was observed in our highest quality crystals.

We have undertaken an annealing program to increase the quality of the crystals. A previous annealing program carried out at Birmingham University, U.K. indicated that the optimal annealing conditions for $Sr_3Ru_2O_7$ were ~1000°C in air for two weeks [19]. Under these conditions, impure crystals (grown under 100% $O_2$) exhibited little change in $\rho_{res}$: $\rho_{res}$ did not fall to below 2 μΩcm in nearly all cases. For high quality crystals (grown under 10%$O_2$+90%Ar atmosphere), $M_{FM}$ decreased by up to two orders of magnitude to as low as $10^{-4}$ emu/g after annealing at 1000°C in air for two weeks. This is intriguing because it suggests that the ferromagnetic phases have been absorbed in to the more stable main $Sr_3Ru_2O_7$ phase under this thermal condition. No study on the effect of annealing on $\rho_{res}$ was made for the new generation of crystals.

**5. Summary and conclusions**

We have succeeded in synthesizing ultra-high-quality single-crystals of $Sr_3Ru_2O_7$ using an infrared image furnace. The crystal quality is now closer to that of $Sr_2RuO_4$ and we are able to produce single crystals of up to 1 g. The high vapour pressure of $RuO_2$ hampers the crystal growth and we used a new approach of measuring the evaporated mass during the growth to determine the atomic ratio of the crystal. This knowledge allowed us to more effectively optimise the growth by predicting the correct nominal atomic ratio for subsequent crystal growths. We also discuss the effect of other growth parameters on the mass lost during the growth. Finally, we have discussed the presence of impurity phases in our crystals. High temperature annealing (1000°C in air for 2 weeks) has also proved to be beneficial to increasing the homogeneity of the $Sr_3Ru_2O_7$ crystals. The improved FFFZ method based on the systematic procedure developed in this study should be useful to other systems for which a high volatility of a constituent material makes the standard TSFZ method inapplicable.

**Acknowledgements**

We acknowledge useful discussions with S. I. Ikeda, N. Kikugawa, H. Fukazawa, and J. C. Hooper. This work has been supported by Grants-in-Aid for Scientific Research from the Japan Society for Promotion of Science (JSPS) and from the Ministry of Education, Culture, Sports, Science, and Technology (MEXT). RSP is supported by a Grant-in-Aid for the 21st Century COE "Center for Diversity and Universality in Physics" from MEXT and is a fellow of JSPS.

|  | $Sr_2RuO_4$ | $Sr_3Ru_2O_7$ | $Sr_3Ru_2O_7$ |
|---|---|---|---|
| $n$ | 1.15 | 1.55 | 1.68 |
| $V_1$ (mm/hr) | 45 | 20 | 15 |
| Atmosphere content | 10%$O_2$+90%Ar | 100%$O_2$ | 10%$O_2$+90%Ar |
| Pressure $P$ (bar) | 3 | 10 | 10 |
| $\rho_{res}$ (μΩcm) | ~0.2 | ~2.0 | 0.25 |
| Mass loss $L$ (%) | 6.0 % | 10.3 % | 10.5 % |
| $n'$ | 0.99 | 1.23 | 1.33 |

Table 1. The growth conditions used for $Sr_2RuO_4$ and $Sr_3Ru_2O_7$ and the analysis results. The growth conditions in the first two columns were taken from refs. [8,15]. Conditions in the third column were developed in the present study.

| $n$ | $O_2/(Ar+O_2)$ | $V_1$ (mm/hr) | $L$ (%) | $n'$ | Batch no. | Diameter (mm) | $T$ (°C) | PC |
|---|---|---|---|---|---|---|---|---|
| 1.60 | 1 | 25 | 7.0 | 1.38 | C668c | 2.4 | 2200 | 214+Ru |
| 1.60 | 10 | 25 | 8.1 | 1.34 | C668a | 2.9 | 2180 | 214+327 |
| 1.60 | 30 | 25 | 10.1 | 1.28 | C668b | 3.2 | 2200 | 214+327 |
| 1.60 | 100 | 25 | 11.2 | 1.25 | C668d | 5.0 | 2180 | 327+214 |

Table 2. The dependence of the mass loss ratio $L$ on the oxygen partial pressure $P_{O2}$. All growths were made at 10 bar pressure. $n=2N(Ru)/N(Sr)$: nominal ratio of the feed rod; $O_2/(Ar+O_2)$: oxygen percentage of the mixed atmosphere $P_{O2}$; $V_1$: crystal growth speed; $L$: the percentage of the mass lost during the growth; $n'$: the resultant ratio of the crystal; $T$: temperature of the floating zone measured using a two-colour pyrometer (Raytek: model MR1 SCCF) with an estimated error of 50°C; PC: Phase constitution of the crystal rod in order of decreasing volume fraction. 214≡$Sr_2RuO_4$; 327≡$Sr_3Ru_2O_7$ and Ru≡ruthenium metal. The PC was determined using powder X-ray diffraction.

| $n$ | $O_2/(Ar+O_2)$ | $V_1$ (mm/hr) | $L$ (%) | $n'$ | Batch no. | PC |
|---|---|---|---|---|---|---|
| 1.68 | 10 | 10 | 13.8 | 1.23 | C642c | 214+327 |
| 1.68 | 10 | 15 | 10.5 | 1.34 | C642a | 327 |
| 1.68 | 10 | 30 | 18.0 | 1.10 | C642b | 327+43_10+214 |

Table 3. Crystal growth of the strontium ruthenates using a variable growth speed $V_1$. All growths were made at 10 bar pressure. $n=2N(Ru)/N(Sr)$: nominal ratio of the feed rod; $O_2/(Ar+O_2)$: oxygen percentage of the mixed atmosphere $P_{O2}$; $V_1$:crystal growth speed; $L$: the percentage of the mass lost during the growth; $n'$: the resultant ratio of the crystal; PC: Phase Constitution of the crystal rod in order of decreasing fraction. 214≡$Sr_2RuO_4$; 327≡$Sr_3Ru_2O_7$ and 43_10≡$Sr_4Ru_3O_{10}$. The PC was determined using powder X-ray diffraction.

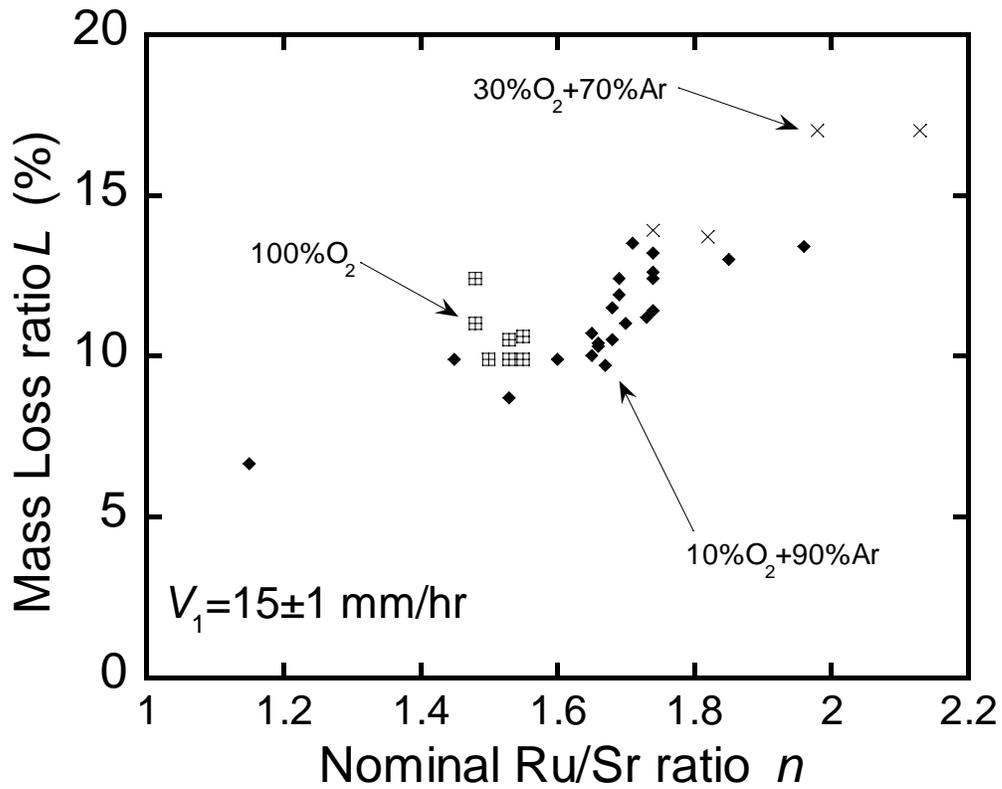

Figure 1. Nominal Ru/Sr ratio $n$, defined as $2N(\text{Ru})/N(\text{Sr})$, versus mass lost during crystal growth of the strontium ruthenates in a floating zone furnace. The crystal growth speed $V_1$ of each growth is between 14 and 16 mm/hr. The total pressure $P$ is 10 bar for each growth.

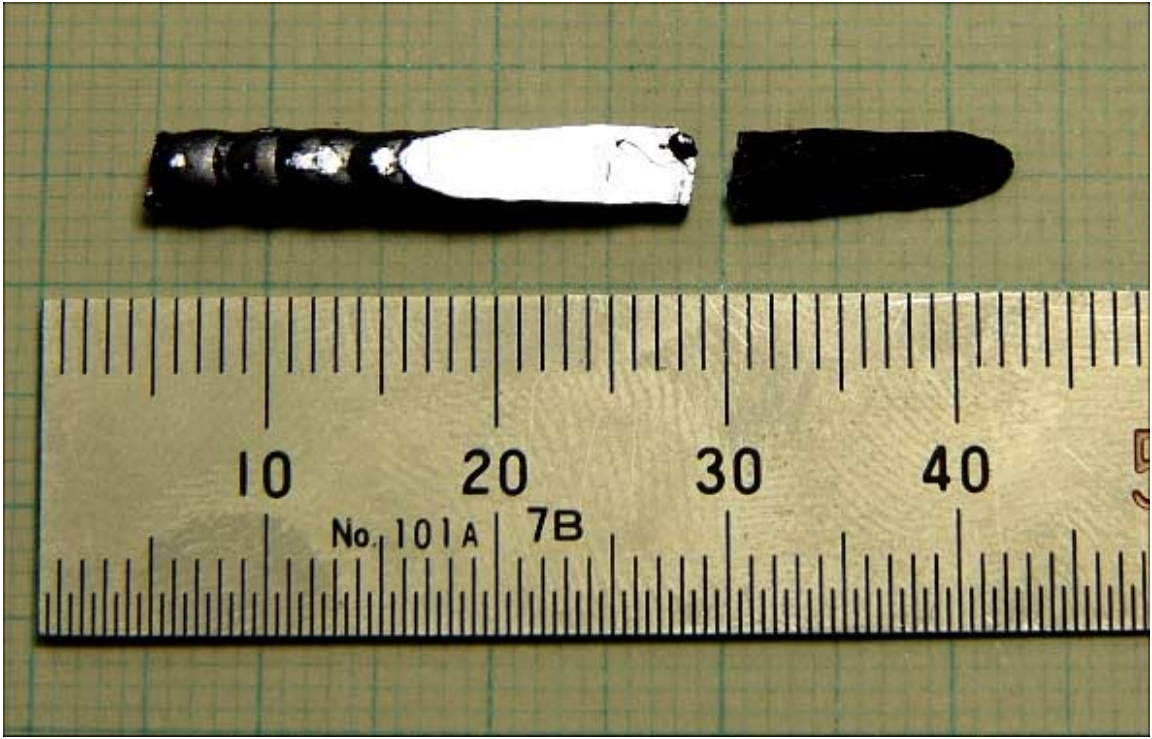

Figure 2. Photograph of two single crystals of $Sr_3Ru_2O_7$ grown under optimised growth conditions. The shiny face in the picture is the *ab* plane. The scale is in mm.

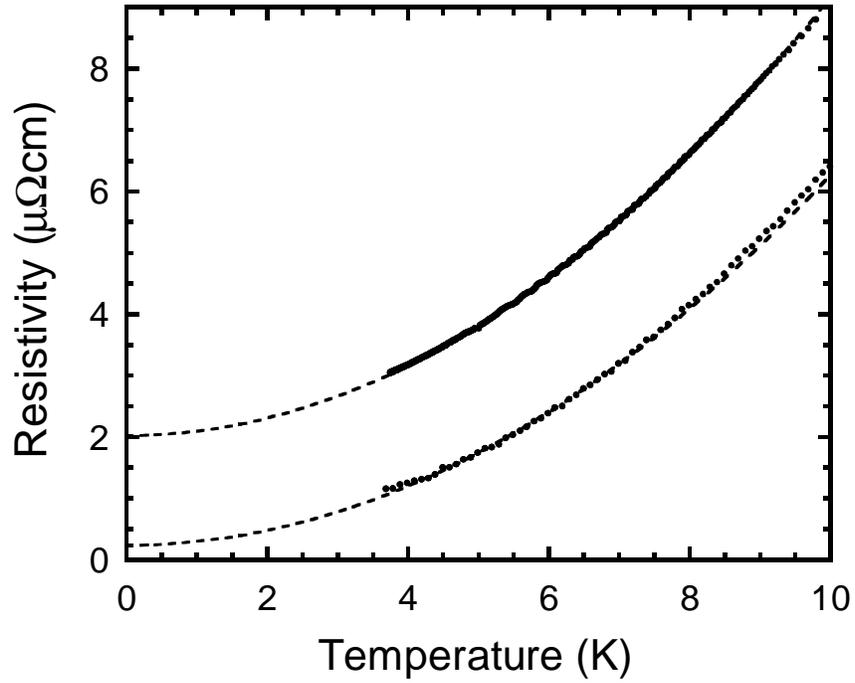

Figure 3. Resistivity versus temperature for two $Sr_3Ru_2O_7$ crystals. The current was 1 mA parallel to the *ab* plane and the dimension of the crystals were typically 3mm×1mm×0.2mm. Electrical connections were made to the crystals using gold wire and high temperature cured (500°C for 5 minutes) silver epoxy (Dupont 6838). The upper curve is a crystal grown under 100% $O_2$ atmosphere and the lower curve is a crystal grown under optimised conditions (10% $O_2$ + 90% Ar). The data are normalised at 300 K to 232 µΩcm. The dotted lines are Fermi liquid $T^2$ fits to the data between 4 K and 8 K.